\documentstyle[aps,pre,twocolumn,amssymb]{revtex}
\begin{document}

\title{\textbf Non-periodic delay mechanism in time-dependent chaotic scattering}
\author{P. K. Papachristou$^1$, F. K. Diakonos$^1$, E. Mavrommatis$^1$ and V. Constantoudis$^2$}
\address{$^1$ Department of Physics, University of Athens, Panepistimiopolis, 157 71 Athens, Greece\\
$^2$ Department of  Physics, National Technical University, Zografou Campus, 157 80 Athens, Greece}
\date{\today}
\maketitle

\begin{abstract}
We study the occurence of delay mechanisms other than periodic orbits in systems with time dependent potentials that exhibit chaotic scattering. By using as model system two harmonically oscillating disks on a plane, we have found the existence of a mechanism not related to the periodic orbits of the system, that delays trajectories in the scattering region. This mechanism creates a fractal-like structure in the scattering functions and can possibly occur in several time-dependent scattering systems.
\end{abstract}

\section{Introduction}
\label{intro}
Chaotic scattering has been an area of intensive research in recent years. The term chaotic scattering is used to indicate the very sensitive dependence of the variables characterizing the outcoming trajectories on the variables characterizing the incoming trajectories \cite{SMI,ECK,JOU,OTT}. Chaotic scattering is usually associated with the appearance of fractal structure in the scattering functions.
\par
Although chaotic scattering has been extensively studied in both classical and quantum mechanics, the vast majority of studies are concerned with systems where the scattering potential does not depend on time. A few studies of time-dependent chaotic scattering systems have recently appeared \cite{GRE,LU,JOS,WIE,MEY,FEN,ALE,ANT,MAT} but there is still a lack of understanding concerning the generic features of such systems.
\par
It is well known that in time-independent scattering systems the trajectories of the scattered particle are delayed in the scattering region whenever the initial conditions of the scattered particle fall onto the stable manifold of one of the infinitely many unstable periodic orbits of the system. This mechanism gives rise to singularities in the delay time functions which are known to have a Cantor-set structure. The situation does not appear to be the same in time-dependent scattering systems. Our aim is to investigate the occurence of delay mechanisms other than periodic orbits in time-dependent systems that exhibit chaotic scattering. We use a model system that is similar to that of Antill\'{o}n \textit{et al.} \cite{ANT}. It consists of two circular disks on a plane that oscillate harmonically with time. The static counterpart of this system is integrable leading to smooth scattering functions \cite{JOS}. A harmonic oscillation law is used in order to avoid the superposition effects of the several harmonics. Our study could serve as a first step towards the deeper understanding of certain scattering processes off oscillating targets, such as molecules, atomic clusters, surfaces and nuclei in the cluster model.
\par
The outline of the paper is as follows: In Sec.~\ref{description} we introduce our model. In Sec.~\ref{results} we present the results of our numerical investigations. In Sec.~\ref{interp} we attempt to give an interpretation of the above results by identifying the mechanism that leads to the delay of the trajectories in the scattering region. In Sec.~\ref{concl} we summarize our main results and comment on the possible extensions of this work.

\section{Description of the model}
\label{description}
We consider the scattering of a free point particle off two circular harmonically oscillating disks on a plane, as shown in Fig. 1. The disks are considered to be much heavier than the scattered particle. The recoil of the disks at the collision is not taken into account and the energy of the system is not conserved. The centers of the disks are oscillating harmonically along the same axis. The position of the center of the i-th disk is given by the equation
\begin{equation}
\textbf{d}_{i}(t)=\textbf{d}_{i}^{(0)}+ \textbf{A}_i\sin(\omega_{i}t+\phi_{i}),
\end{equation}
where $\textbf{d}_{i}^{(0)}$ denotes the equilibrium position of the center, $\textbf{A}_i$ is a vector directed along the axis of oscillation having magnitude equal to the amplitude of the oscillation, $\omega_i$ is the angular frequency of the oscillation and $\phi_i$ is the initial phase. The position $\textbf{r}(t)$ and the velocity $\textbf{u}(t)$ of the particle in the time interval between the $(n-1)$-th and the $n$-th bounce are given by
\begin{eqnarray}
\textbf{r}(t)=\textbf{r}_{n-1}+(t-t_{n-1})\textbf{u}_{n-1} \nonumber \\
\textbf{u}(t)=\textbf{u}_{n-1},
\label{eq:betw}
\end{eqnarray}
where $\textbf{r}_{n-1}$ denotes the position of the (n-1)-th bounce, $\textbf{u}_{n-1}$ denotes the velocity of the scattered particle after the $(n-1)$-th bounce and $t_{n-1}$ denotes the time when that bounce occured. In order to find the time $t_n$ when the next bounce will occur, we must solve the equation
\begin{equation}
\left\| \textbf{d}_{i}(t)-\textbf{r}(t)\right\|^2=R^2_i,
\label{eq:cond}
\end{equation}
where $R_i$ denotes the radius of the i-th disk. This equation must be solved twice (for $i=1,2$) and the smallest non-negative solution has to be kept. The condition~(\ref{eq:cond}) leads to the equation
\begin{eqnarray}
c_{1}(t-t_{n-1})^2+c_{2}\sin^2(\omega_{i}t+\phi_i) \nonumber \\
+c_{3}(t-t_{n-1})\sin(\omega_{i}t+\phi_i)+c_{4}(t-t_{n-1}) \\ \label{eq:eqn}
+c_5\sin(\omega_{i}t+\phi_i)+c_6=0 \nonumber
\end{eqnarray}
where the coefficients $c_1,\ldots,c_6$ are given by
\begin{eqnarray}
c_1=\textbf{u}_{n-1}^2 \nonumber \\
c_2=\textbf{A}_{i}^2 \nonumber \\
c_3=-2\textbf{u}_{n-1}\cdot\textbf{A}_i \nonumber \\
c_4=2(\textbf{r}_{n-1}-\textbf{d}_{i}^{(0)})\cdot\textbf{u}_{n-1}\\
c_5=2(\textbf{r}_{n-1}-\textbf{d}_{i}^{(0)})\cdot\textbf{A}_{i} \nonumber \\
c_6=(\textbf{r}_{n-1}-\textbf{d}_{i}^{(0)})^{2}-R_{i}^2. \nonumber
\end{eqnarray}
Equation~(4) is solved numerically to obtain $t_n$. Using equation~(\ref{eq:betw}), the position of the next bounce is then given by
\begin{equation}
\textbf{r}_{n}=\textbf{r}_{n-1}+(t_{n}-t_{n-1})\textbf{u}_{n-1}
\end{equation}
Since the mass of the disks is considered to be much larger than that of the scattered particle, the velocity of the particle after the n-th bounce is given by
\begin{equation}
\textbf{u}_{n}=\textbf{u}_{n-1}-2[\textbf{n}\cdot(\textbf{u}_{n-1}-\textbf{V}_i)]\cdot\textbf{n}
\end{equation}
where $\textbf{n}$ is the normal to the disk at the point of the impact and $\textbf{V}_i$ is the velocity of the involved disk at the instant of the collision, which is given by
\begin{equation}
\textbf{V}_i=\textbf{A}_{i}\omega_{i}\cos(\omega_{i}t_{n}+\phi_i).
\end{equation}
In our study we have chosen the x-axis as the axis of oscillation ($\textbf{A}_{i}=(A_i,0)$). We have also chosen the radii and the angular frequencies of the disks to be the same ($R_1=R_2=R$, $\omega_1=\omega_2=\omega$).
By following the above procedure, we iterate numerically the map $(\textbf{r}_n,\textbf{u}_n,t_n)\rightarrow(\textbf{r}_{n+1},\textbf{u}_{n+1},t_{n+1})$ and obtain the trajectories for a large number of initial conditions. 
\par

\section{Numerical results}
\label{results}
\subsection{Fractal-like scattering functions}
The scattering region is defined as a circular domain of radius $R_0>R_i$ centered at the origin. As initial conditions we choose $\textbf{r}(0)=0$  and $\textbf{u}(0)=(u_{0}\cos\alpha,u_{0}\sin\alpha)$. For the majority of our numerical calculations we have chosen the parameters of the system to be $R=1$, $\textbf{d}_1^{(0)}=-\textbf{d}_2^{(0)}=(-3/2,0)$, $A_1=A_2=A=0.1$, $\phi_1=\pi$, $\phi_2=0$, $\omega=10$, $\alpha=10^{-2}rad$ and $R_0=10$. For these values of the parameters the scattered particle exhibits a small number of bounces (typically 4-7), in contrast to Ref.~\cite{ANT} where the number of bounces is much larger (typically more than 100) due to the different choise of the parameters. We have obtained the delay time $T$ that the particle spends in the scattering region, the scattering angle $\phi_{out}$ and the outgoing velocity $u_{out}$ as a function of $u_0$. The results are shown in Fig. 2. A rich fractal-like structure is observed in these scattering functions for rather small values of $u_0$. As $u_0$ increases, the scattering functions become more regular. The dependence of $T(u_0)$ on the angular frequency $\omega$ has also been studied. The results for $\omega=100$ and $\omega=1000$ are shown in Fig. 3. From this figure it is obvious that the change of $\omega$ simply introduces a scaling in the $T(u_0)$ function. In order to study the dependence of the structure of this function on the angle $\alpha$, we have calculated $T(u_0)$ for $\alpha=10^{-3}rad$ and $\alpha=10^{-1}rad$. The results are shown in Fig. 4. We observe that as the angle increases the function becomes more regular in the sense that the range of $u_0$ in which $T(u_0)$ exhibits wild oscillations becomes smaller. In order to investigate the fractal structure of the system, we make successive magnifications of the $T(u_0)$ plot of Fig. 2(a) in a region of $u_0$ around the value 0.8. The results are shown in Fig. 5. We observe that the fractal-like structure breaks at a very small scale. In this $u_0$ region, this breaking scale is found to be around $10^{-7}$. In the following, a more in depth analysis of these findings will be given.

\subsection{Investigation of the fractal-like structure}
\label{sec3.2}
In order to quantify the observations concerning the fractality of the system made in the previous subsection, we have calculated the uncertainty dimension of $T(u_0)$ \cite{LAU,CON}.  This quantity is given by $d_u=1-\beta$, where $\beta$ appears as $f(\epsilon)\approx \epsilon^\beta$. $f(\epsilon)$ is the fraction of uncertain points, for a given value of uncertainty, $\epsilon$, and for randomly chosen points $u_0$. Each $u_0$ is considered to be uncertain if we find that the difference $\left| T(u_0)-T(u_0+\epsilon) \right|$ is larger than a number of order 1. In our calculation we have included as many random points as necessary to obtain 150 uncertain points per run. We have calculated the uncertainty dimension for several values of $\alpha$ and $\omega$ and in all cases $\log f(\epsilon)$ as a function of $\log\epsilon$ can be well fitted by a straight line. The results are shown in Table 1. We can make the following observations:
\begin{enumerate}
\renewcommand{\labelenumi}{\roman{enumi})}
\item The dimension $d_u$ decreases with increasing angular frequency. This was expected since at the $\omega \to 0$ limit the system becomes regular.
\item The dimension $d_u$ decreases with increasing $\alpha$. This was also expected since for smaller values of $\alpha$ the scattered particle senses more strongly the dynamics in the scattering region.
\end{enumerate}
For each angle $\alpha$, there is a scale below which it is not possible to find any uncertain points. This supports our observation that self-similarity does not persist at arbitrarily small scales.

\subsection{Decay law}
We have also investigated the behaviour of the function $N(t)/N_0$ which gives the fraction of particles that remain in the scattering region after time $t$. It is known that for hyperbolic systems $N(t)/N_0$ decays exponentially whereas for systems with marginal orbits and KAM tori it usually obeys a power law \cite{GAS}. In order to stydy numerically the behaviour of $N(t)/N_0$ we need to perform a Monte Carlo simulation by iterating a large number $N_0$ of randomly selected initial conditions. In our system, in contrast to the static 2-disk system, an averaging over the initial velocity of the particle has also to be performed since the system is not conservative. In our calculation for the oscillating system, $u_0$ is uniformly distributed in $(0,2]$ since this is the range of $u_0$ where all the peaks of $T(u_0)$ occur. The angle $\alpha$ has been chosen in $(0,0.7]$ since for values of $\alpha$ in this range the particle exhibits at least one collision before exiting the scattering region. An averaging over the initial phases has also been performed. We have used $10^7$ orbits. In order not to overemphasize a special class of peaks we choose $R_0=2.5$ for our calculations. The result is shown in Fig. 6. From this figure it is clear that for our system $N(t)/N_0$ is very close to a power law with an exponent which for the chosen values of the parameters is found to be approximately equal to $-2.38$. The corresponding static system is expected to have an exponential decay law. In this case, no averaging in $u_0$ is performed since the system is conservative. If we average in $u_0$ for the static system we will also obtain a power law with an exponent equal to $-2$. If, on the other hand, we exclude from our calculations the "uncertain" trajectories (for which $|T(u_0,\phi,\alpha)-T(u_0+\epsilon,\phi,\alpha)|$ is larger than a number of order 1) for the oscillating system we find a decay which for large times follows approximately a power law with an exponent equal to $-5.09$. The result is shown in Fig. 6. From the above it becomes clear that although the presence of the oscillation accelerates the escape of the particles, the presence of the high peaks in the $T(u_0)$ function (uncertain points) introduces a delay of the particles and slows down the escape. The origin of these peaks will be discussed in the following section.

\par

\section{Interpretation of the results: a new delay mechanism}
\label{interp}

By analyzing the trajectories that stay for long times in the scattering region, we found that these do not exhibit a large number of bounces. The observed delay comes from orbits along which the scattered particle loses much of its energy and therefore traverses segments of its orbit with a very low velocity. This sudden loss of energy can happen at any of the bounces provided that $\|\textbf{u}_{n-1}-2\textbf{V}_i\|$ is small but not as small as for the particle to bounce on the same disk again. We have classified the peaks of the $T(u_0)$ plot according to the bounce which leads to the major loss of energy. The result is shown in Fig. 7. Prominent peaks that are due to the first bounce are observed. We also observe that around each of the peaks there is a rich fractal-like structure. In the following we are going to give a qualitative interpretation of the above observations by using a simple one-dimensional model which neglects the curvature of the disks. This model can yield quantitative results only for the first and the second bounce, since with the parameters chosen, the scattered particle senses strongly the curvature of the disks after the second bounce.

Setting $\alpha=0$ the dynamics is limited to one spatial dimension. In this case the system resembles to the Fermi acceleration model \cite{BRA,LIC}. The particle starts at the origin with initial velocity $\textbf{u}(0)=(u_0,0)$. In order to find the time $t_c$ when the first collision with the right disk occurs, we have to solve the equation
\begin{equation}
u_0t={D \over 2}-R+A\sin (\omega t).
\end{equation}
The time $t_c$ can be thought of as the abscissa of the first point of intersection of the straight line $u_0t$ with the sinusoidal curve ${D \over 2}-R+A\sin (\omega t)$ (see Fig. 8). From this figure it is obvious that $t_c$ is a discontinuous function of $u_0$. Discontinuities occur for the values of $u_0$ for which the line is tangent to the sinusoidal curve. If we denote as $t_c^+$ the value of $t_c$ after the discontinuity, the value of $u_0=u_0^*$ at which the discontinuity occurs is given by
\begin{equation}
u_0^*=A \omega \cos(\omega t_c^+).
\end{equation}
Therefore, the velocity of the particle after the first collision $u_1$ is also discontinuous as a function of $u_0$. At this point we should stress that the oscillation law need not to be only harmonic for these discontinuities to occur.  In Fig. 9(a) a plot of $u_1$ as a function of $u_0$ is shown. We observe sharp peaks that get denser as $u_0 \to 0$. For $u_0 \to 0$, the initial velocity that corresponds to a given $t_c$ is given by
\begin{equation}
u_0\cong {{R-{D \over 2}-A} \over {t_c}}.
\end{equation}
Therefore, the distance between two successive peaks on the $u_0$ axis is given by
\begin{equation}
\Delta u_0\cong {{R-{D \over 2}-A} \over {t_c^2}}\Delta t_c
\end{equation}
where $\Delta t_c\cong {2\pi \over \omega}$. Combining the above two relations, we conclude that $\Delta u_0\sim u_0^2$ as $u_0 \to 0$. The density of the peaks therefore increases as ${1 \over u_0^2}$ as $u_0 \to 0$.

We observe that $u_1$ gets close to 0 near the discontinuities ($u_0\cong 2V$, where $V$ is the velocity of the disk involved in the collision). If the velocity after the collision is small, but not small enough as for the particle to rebounce on the same disk, there is a delay of the particle between the first and the second bounce. We expect this delay to be present in the original system ($\alpha\ne0$) and to manifest itself as a peak in the $T(u_0)$ plot, since for small $\alpha$ the curvature of the disks can be neglected for the first collision.

In the following we will give a qualitative interpretation of the fact that there is a rich fractal-like structure around the peaks of $T(u_0)$ for the original system . The quantity $\left| {1/u_n} \right|$ is a measure of how much time is spent between the $n$-th and $(n+1)$-th collision for the one dimensional system. In Fig. 9(b) a plot of $\left| {1/u_1} \right|$ as a function of $u_0$ for the one-dimensional system is shown. We observe that the peaks of this plot are very close to the peaks of the $T(u_0)$ plot that correspond to the delay of the particle between the first and the second bounce. The $T(u_0)$ plot can be thought of as several iterations of $\left| {1 /u_n} \right|$ plots which are expected to have a structure similar to the $\left| {1/u_1} \right|$ plot. For $u_0$ that corresponds to a peak of the $T(u_0)$ plot, $u_1$ falls in the low velocity region, where the structure of the $\left| {1/u_1} \right|$ plot is very dense. We therefore expect fingerprints of this dense structure to be apparent around any $u_0$ value which maps onto the low velocity region after some bounce. Since there is a lower bound on the velocity with which the particle can leave a disk, we do not expect the structure around the peaks to be infinitely dense. This is consistent with our observation that the fractal structure of $T(u_0)$ breaks at some small $u_0$ scale. The higher a peak of the $T(u_0)$ plot is, the smaller the velocity of the particle along the orbit and the lower the breaking scale. The above mechanism also explains the presence of peaks in the scattering functions of  Ref.~\cite{ANT}, however the structure there is more dense due to the different choise of the parameters. Furthermore, the mechanism is also present when the oscillating hard disks are replaced by two oscillating potential hills. A detailed analysis of the dynamics of such a system is left for a future study.

\section{Conclusions and outlook}
\label{concl}
In the present paper we have studied the effects of time-dependence on the scattering proccess of a point particle off two harmonically oscillating hard disks. A new mechanism leading to long-lived scattering trajectories has been found. It is associated with the energy loss of the scattered particle at the collisions. This mechanism is not related to the periodic orbits of the system and induces a fractal-like structure in the scattering functions. At the statistical level, the mechanism manifests itself as a change in the properties of the fraction $N(t)/N_0$ of particles that remain in the scattering region after time t. Although this function still obeys a power law, the absolute value of the corresponding exponent is reduced by a significant amount. An interesting question that will be studied in the future is how time-dependence affects the dynamics of a system whose static counterpart is chaotic, such as the three disk system \cite{GAS-RIC}. As an extension to this work, the transport properties of a lattice gas consisting of oscillating disks will be studied. Another open question is the quantum manifestation of the new mechanism.

\section{Ackowledgments}
We thank Prof. L. Cederbaum and Dr. P. Schmelcher for helpful discussions. P. P. acknowledges a scholarship from the Greek State Scholarships foundation (IKY).

\newpage
\begin{table}
\begin{tabular}{lrcccc}
               & $\omega$(a.u.) &   $1$ &        $10$ &   $100 $\\
$\alpha$(rad)  &&     &           &       \\
\hline
$10^{-3}$ & & $0.63\pm 0.01$     & $0.62\pm 0.01$ &  $0.55\pm 0.02$ \\
$10^{-2}$ & & $0.61\pm 0.02$      & $0.58\pm 0.01$ & $0.54\pm 0.01$ \\
$10^{-1}$ & & $0.58\pm 0.01$      & $0.53\pm 0.02$ &  $0.52\pm 0.01$ \\
\end{tabular}
\vspace{10pt}
\caption{The uncertainty dimension of the system for various values of the initial angle $\alpha$ and the angular frequency $\omega$.}
\end{table}

\begin{figure}
\caption{The two oscillating disks on the $x-y$ plane. The positions of disks at time $t_{n-1}$ are drawn with the dotted line and at $t_n$ with the solid line. The point A, which is defined with the vector $\textbf{r}_{n-1}$, is the point of the $(n-1)$-th bounce. The point B, which is defined by $\textbf{r}_n$, is the point of the $n$-th bounce. The dashed line with the arrows represents a segment of a trajectory that bounces between the two disks.}
\end{figure}

\begin{figure}
\caption{(a) The delay time $T$, (b) the scattering angle $\phi_{out}$ and (c) the outgoing velocity $u_{out}$ as a function of the initial velocity $u_0$ for angular frequency $\omega=10$ and initial angle $\alpha=10^{-2}rad$. All the plots were created by iterating $10^4$ initial conditions.}
\end{figure}

\begin{figure}
\caption{The delay time $T$ as a function of the initial velocity $u_0$ for $\alpha=10^{-2}rad$ and $\omega=100$ (a), $\omega=1000$ (b).}
\end{figure}

\begin{figure}
\caption{The delay time $T$ as a function of the initial velocity $u_0$ for $\omega=10$ and $\alpha=10^{-3}rad$ (a), $\alpha=10^{-1}rad$ (b).}
\end{figure}

\begin{figure}
\caption{Magnifications of Fig. 2(a) around the value 0.8 of the initial velocity $u_0$. Although in (a) a self similar structure appears to exist, it is found that this structure does not persist at arbitrarily small scales as shown in (b) and (c).}
\end{figure}

\begin{figure}
\caption{The fraction $N(t)/N_0$ of particles remaining in the scattering region at time $t$. The thick solid curve includes all the trajectories and the thin solid curve includes only the "certain" trajectories (see Sec.\ref{sec3.2}). For large $t$ both curves approximately obey power laws with different exponents. The dashed lines show the corresponding linear fits. The plots were created by iterating $N_0=10^7$ initial conditions.}
\end{figure}

\begin{figure}
\caption{Some of the major peaks of Fig. 2(a) are classified according to the bounce that leads to the major energy loss of the scattered particle. The symbols $\bigtriangleup$,$\Box$ and $\circ$ denote the first, the second and the third bounce respectively.}
\end{figure}

\begin{figure}
\caption{The abscissa of the point of intersection of the sinusoidal curve $D/2-R+\sin\omega t$ and the straight line $u_0t$ is the time $t_c$ when the first bounce occurs for the one-dimensional system ($\alpha=0$). The slope of the straight line is $u_0$. For values of $u_0$ for which the straight line is tangent to the sinusoidal curve, $t_c$ is discontinuous as a function of $u_0$. The value of $t_c$ after the discontinuity is denoted as $t_c^+$.}
\end{figure}

\begin{figure}
\caption{(a) The velocity $u_1$ after the first bounce as a function of the initial velocity $u_0$ for the one-dimensional system ($\alpha=0$). (b) $\left|1/u_1\right|$ as a function of $u_0$ for the one-dimensional system. The locations of the singularities are very close to the locations of the peaks of Fig. 7 that correspond to a loss of energy of the scattered particle at the first bounce.}
\end{figure}

\end{document}